\documentclass[12pt]{article}

\usepackage{lineno,hyperref}
\modulolinenumbers[5]

\usepackage{amsmath}
\usepackage{amssymb}
\usepackage{graphicx}
\usepackage{color, colortbl}
\usepackage{multirow}
\usepackage{xcolor}
\usepackage{arydshln}
\usepackage{gensymb}
\usepackage{tikz}
\usetikzlibrary{fadings, shadings}
\usepackage{dashrule}
\usepackage{harvard}
\input{macrostrans.tex}

\begin{document}

\title{Numerical optimization of a prestressed auxetic metamaterial for vibration isolation\footnote{This is the authors' original version of a paper submitted for publication in 2019, June.}}

\author{\medskip
\textbf{A Pyskir\textsuperscript{1,2}, M Collet\textsuperscript{1}, Z Dimitrijevic\textsuperscript{2}, and C H Lamarque\textsuperscript{3}} \\
\textsuperscript{1} LTDS UMR-CNRS 5513,\\
\'Ecole Centrale de Lyon,\\
36 avenue Guy de Collongue, 69134 \'Ecully, France\\
\medskip
e-mail: \textbf{adrien.pyskir@ec-lyon.fr}\\
\textsuperscript{2} PSA Groupe, DRIA/DSTF/SMMS,\\
\medskip
Route de Gisy, 78140 V\'elizy-Villacoublay, France\\
\textsuperscript{3} LTDS UMR-CNRS 5513,\\
\'Ecole Nationale des Travaux Publics de l'\'Etat,\\
 3 Rue Maurice Audin, 69518 Vaulx-en-Velin, France}

\maketitle

		
\begin{abstract}
We present a numerical study on an enhanced periodic auxetic metamaterial. Rotating squares mechanism allied to precompression induced buckling give these elastic structures exotic properties. The static properties of the reference structure and the enhanced ones are first compared. After numerical analysis to ascertain the differences between several band calculation methods, we demonstrate the effect of precompression issued stress field on the dispersion diagram of the metamaterial. An optimization study is then performed to assess the potential vibration isolation improvement obtained with the new design. As a result, the bandgaps widths and range are found to be greatly increased by the geometric modifications proposed.
\end{abstract}

\vspace{2pc}
\noindent{\it Keywords}: Elastic metamaterial, vibration isolation, auxetic, bandgap



\section{Introduction}
For more than two decades, interest for metamaterials has been steadily increasing, with inspiring applications like invisibility cloaks \cite{SCH06} or perfect lens \cite{PEN00}. Their multidisciplinary aspect contributes greatly to the growing effort to improve these usually periodic structures, exhibiting properties unseen in conventional materials \cite{KAD13}. If lots of concepts found in literature use resonance phenomena to achieve these properties, other types of systems can also display exotic properties. As an example, one can think of pentamode metamaterials, solid structures designed to achieve virtually zero shear modulus, just as liquids behave \cite{BUC14}.\\

Another class of often studied structures is that of auxetic metamaterials, characterized by a negative Poisson's ratio. It means that they expand in one or several transverse directions under axial traction, and tend to transversally contract when axially compressed. Though such architectures have been studied for quite a long time \cite{ROB85,GIB82}, at first only their static properties, like Poisson's ratio and effective stiffness modulus, were described. As for their dynamic characteristics, they gained attraction at the end of the nineties \cite{SCA00}. Like many other metamaterials, they can feature frequency bands where no waves can propagate, called bandgaps. For obvious reasons, bandgaps are extremely interesting for vibration isolation and to find bandgaps as wide and low frequency possible with compact solution is an overriding goal in this study.\\
Among these structures, some harness buckling to exhibit peculiar static and dynamic properties, with potential applications in vibration isolation \cite{BER08,SHI15}.\\

This paper focuses on 2D numerical simulation of such structures. The computed geometries are introduced in a first part, followed by a description of the numerical methods used in this study. The next part deals with the static compression computations of the geometries, and gives base results for the dynamic study, detailed in the last part. In particular, effects of strain and stress fields from a static load on wave properties are analysed. As we aim to develop structures with the best vibration isolation properties possible, a geometric optimization with several criteria is finally proposed and discussed.\\


\section{Design}
\begin{figure}[ht!]
	\centering
	\includegraphics[width=12cm]{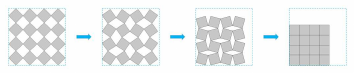}
	\caption{Rotating square base mechanism. Configurations for different angles between touching squares(45°, 30°, 15°, 0°)}
	\label{fig:geom_theo}
\end{figure}

The herein studied geometry is derived from the rotating squares structure, a quite classical design used by numerous authors for two decades \cite{GRI00}. The auxetic mechanism behind this structure is a simple geometric one, as it consists in 2D polygons, each one rotating with respect to its neighbours, and thus changing the global size of the system in both directions (See figure \ref{fig:geom_theo}). 

\begin{figure}[ht!]
	\centering
	a.\includegraphics[height=0.12\textheight]{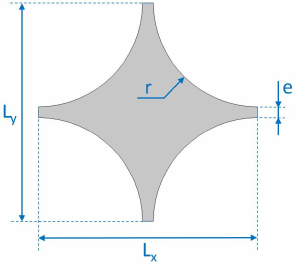}
	\includegraphics[trim=0cm 0cm 0cm 0.5cm, clip=true, height=0.05\textheight]{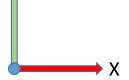}
	b.\includegraphics[height=0.12\textheight]{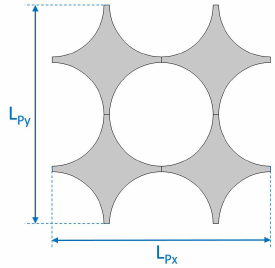}
	\caption{Initial structure $S_0$: a) unit cell and b) periodic pattern.}
	\label{fig:cell}
\end{figure}

However, this type of design is hard to implement practically due to point contacts between polygons. Even though some papers use systems really close to it \cite{DEN17}, many of them choose a simple modification solving the point contact problem and the resulting stress concentration near sharp angles. To do so, the square holes - on the first configuration in figure \ref{fig:geom_theo}) - between polygons can be replaced with round holes. The corresponding geometry is shown in figure \ref{fig:cell}.\\

For all steps and geometries in this study, the same initial geometrical parameters will be kept as follow: $L=L_{x,0}=L_{y,0}=10mm$, $e=1mm$. The radius $r$ in 
figure \ref{fig:cell}a is directly derived from previous parameters through $r=(L-e)/2$. Moreover all the results in this paper are computed for a thickness $h_z=40mm$. It should also be noted that the cell dimensions $L_x$ and $L_y$ are bound to vary with the structure deformation $\varepsilon=(L_{y,0}-L_{y})/{L_{Py}}$, and are not equal anymore - though very close - as soon as $\varepsilon>0$.\\ 

\begin{figure}[ht!]
	\centering
	\includegraphics[width=12cm]{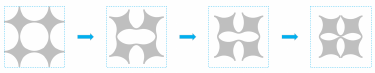}
	\caption{Mechanical behaviour during compression of periodic pattern. Configurations for different values of deformation ($0\%$, $10\%$, $20\%$, $\approx30\%$).}
	\label{fig:geom_phys}
\end{figure}

Whereas this geometry keeps the auxetic properties of the original one, their respective mechanical behaviour strongly differs. Instead of simply rotating and smoothly closing as described by figure \ref{fig:geom_theo}, the rotation of the polygons now triggers the buckling of the intermediate thin ligaments. As a result, the contact between cells happen without total closing of the holes (See figure \ref{fig:geom_phys}).\\

\begin{figure}[ht!]
	\centering
	a.\includegraphics[height=0.12\textheight]{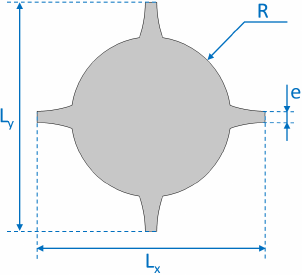}
	\quad
	b.\includegraphics[height=0.12\textheight]{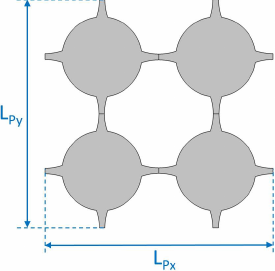}
	\quad
	c.\includegraphics[trim=1.5cm 0.6cm 1.5cm 1.5cm, clip=true, height=0.12\textheight]{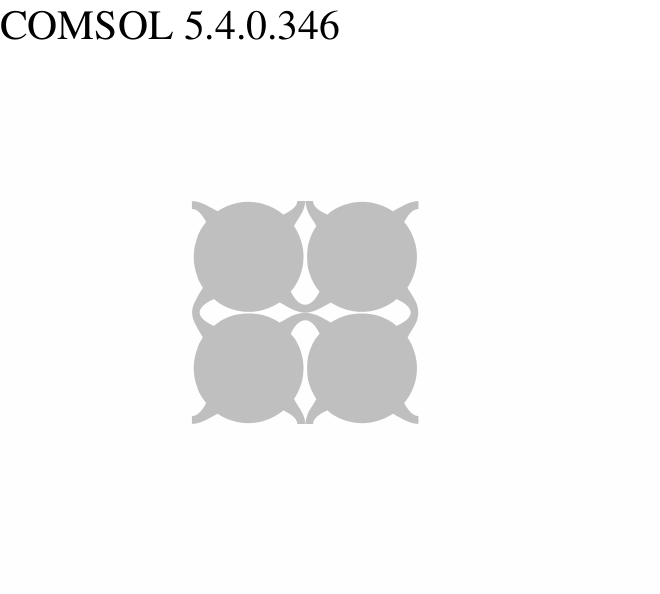}
	\caption{Example of modified configuration with its characteristic dimensions for $R=3.5mm$: a. unit cell, b. periodic pattern, and c. pattern at $\approx30\%$ deformation level.}
	\label{fig:cell_enh}
\end{figure}

This finding led to a new design, making use of the available space without modifying the static behaviour of the structure. The concept is described in figure \ref{fig:cell_enh}. It can be seen as a disk of matter centered on the middle point of initial design $S_0$.

It should be noted that the unit cell shown in figure \ref{fig:cell}a is not periodic anymore after buckling. Instead, four cells should be considered to obtain a periodic pattern. The characteristic lengths of a pattern along $X$ and $Y$ are then respectively $L_{Px}=2L_{x}$ and $L_{Py}=2L_{y}$, as explained by Figs. \ref{fig:cell}b and \ref{fig:cell_enh}b.


\section{Numerical methods}\label{s:methodes}
\subsection{Brillouin zone}\label{ss:BZ}
For periodic structures, the Brillouin Zone (or \textit{BZ}) corresponds to the periodic pattern transposed in the reciprocal space, that is to say the wavenumber space \cite{BRI46}. As every pattern in the structure is identical, all the eigenmodes can be obtained from the BZ data only. Even better, Brillouin showed that symmetries in the BZ can be used to further reduce the calculation domain and to find a minimal zone called Irreducible Brillouin Zone (or \textit{IBZ}).  
 
	\begin{figure}[ht!]
	\centering
	\includegraphics[trim=0cm 0cm 0cm 0cm, clip=true,width=3.6cm]{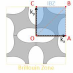}
	\caption{BZ and IBZ depiction for a 2D pattern}
	\label{fig:IBZ}
\end{figure}					

In figure \ref{fig:IBZ}, IBZ is seen to amount to a quarter of the BZ area. In undeformed configuration, this IBZ could even be reduced by half ($OAB$ area only), but as hinted before, as soon as $\varepsilon>0$, the symmetry along $OB$ axis is not true anymore.\\

In the end, the IBZ allows to determine the dynamic behaviour of the whole structure with an optimized computational cost, as the waves in the IBZ and in the structure are considered to be the same. Further simplification is achieved by computing only the IBZ contour - that is to say OA, AB, BC, and OC - visible in figure \ref{fig:IBZ}. The modes are then usually presented in graphs plotting the frequency as function of the wavenumber, call band diagrams or dispersion diagrams. This type of graph is particularly useful to find bandgaps, and thus the isolating performance of the metamaterial.\\

Whereas most articles in litterature compute only the IBZ contour, it is interesting to note that other papers found that depending on the shape of the IBZ, extrema can exist inside the domain \cite{HAR07,MAU18}. Thus if preliminary detection of bandgaps can be achieved through the IBZ contour, validating the exact bandgaps frequency range will require a full sweep of the IBZ. This study implements only five branches which bounds are defined as follow:

\begin{description}
\item[OA:] $k_x\in\Delta\degree/{L_x}$ and $k_y=0$
\item[AB:] $k_x={\pi}/{L_x}$ and $k_y\in{\Delta\degree}/{L_y}$ 
\item[BC:] $k_x\in{\Delta\degree}/{L_x}$ and $k_y={\pi}/{L_y}$
\item[OC:] $k_x=0$ and $k_y\in{\Delta\degree}/{L_y}$
\item[OB:] $k_x\in{\Delta\degree}/{L_x}$ and $k_x\in{\Delta\degree}/{L_y}$
\end{description}
where $\Delta\degree$ is the sweep range $[0,\pi]$.

\subsection{Wave Finite Element Method}\label{ss:WFEM}
Wave Finite Element Method (or \textit{WFEM}) is classical method to compute the dispersion diagram of periodic structures. It is an hybrid method combining analytical approach to finite elements in order to model complex structures with limited computational cost.\\
The method uses the Floquet-Bloch theorem which states that displacements on the boundaries of an \textit{L}-periodic pattern\footnote{periodic pattern where \textit{L} is the space period} are governed by the relation:

\begin{equation}
\mathbf{u}(\omega,x)=\mathbf{\tilde{u}}(\omega,x)exp(\mathrm{i} kx)
\end{equation}
where $\mathbf{\tilde{u}}$ is \textit{L}-periodic.\\
When applied to vibration propagation in a 2D pattern periodic along $X$ and $Y$, which size is $L_{x}\times L_{y}$, and for boundaries only, the equation gives:
 
\begin{equation}
  \left\{
      \begin{aligned}
		\mathbf{u_\mathrm{R}}=\mathbf{u_\mathrm{L}}exp(\mathrm{i} k_{x}L_{x})\\
		\mathbf{u_\mathrm{T}}=\mathbf{u_\mathrm{B}}exp(\mathrm{i} k_{y}L_{y})
      \end{aligned}
    \right.
    \label{eq:Floquet}
\end{equation}
where $\mathbf{u_\mathrm{\bullet}}=[u_{\mathrm{\bullet},x} u_{\mathrm{\bullet},y} u_{\mathrm{\bullet},z}]^t$ is the displacement vector on boundary $\mathrm{\bullet\in\{L, R, T, B\}}$, corresponding respectively to left, right, top, and bottom boundaries. $k_{x}$ (resp. $k_{y}$) is the wavenumber along $X$ (resp. $Y$) axis.\\
The dynamic equilibrium is then defined by the equation:

\begin{gather}
	\mathbf{D}\mathbf{u}=\mathbf{F}\label{eq:edp_dyn}\\
	\mathbf{D}=-\omega^2\mathbf{M}+(1+\mathrm{i}\eta)\mathbf{K}\nonumber
\end{gather}

with $\mathbf{M}$ and $\mathbf{K}$ the respective matrices of mass, and stiffness, $\mathbf{D}$ the dynamic stiffness matrix, $\omega$ the system pulsation, $\eta$ the damping coefficient, and $\mathbf{u}$ and $\mathbf{F}$ the respective vectors of nodal displacements and loads. The characteristic dimension of these matrices and vectors is the number of degrees of freedom (DOFs).\\

By applying the Guyan reduction - which differenciates boundary and internal elements - to (\ref{eq:Floquet}), the following eigenvalue problem is obtained:

\begin{equation}
\left[\mathbf{S}-\lambda_j \mathbf{I_{2n}}\right]\mathbf{\Phi_j}=0 \quad \quad \lambda_j=exp[\mathrm{i}(k_{x,j}L_x+k_{y,j}L_y)]
\label{eq:WFE}
\end{equation}
where $\mathbf{S}$ depends on $\mathbf{D}$ matrix coefficients with differenciated DOFs. $\mathbf{\Phi}=\left(
      \begin{aligned}
		\mathbf{u_\mathrm{L}}\\
		\mathbf{-F_\mathrm{L}}
      \end{aligned}
    \right)$, $\mathbf{I_{2n}}$ is the identity matrix of size $2n$, where $n$ is the number of DOFs.\\
    
By solving (\ref{eq:edp_dyn}) for Floquet periodic solution (See (\ref{eq:Floquet})), one can derive the dispersion diagram of the structure. For further information about the WFEM, quite a number of articles explain the method in details and can be referred to \cite{MAC08,ZHO15}.\\

\subsection{Shifted Cell Operator Method}\label{ss:SC}

For the Shifted Cell Operator Method (or \textit{SCOM}), the partial derivative equation (\ref{eq:edp_dyn}) is rewritten by directly shifting the wavenumbers in the derivative space. It gives:

\begin{equation}
\left[(\mathbf{K}-\omega_j^2\mathbf{M})+\lambda_j\mathbf{L}-\lambda_j^2\mathbf{H}\right] {\mathbf{\Phi_j}}=\mathbf{0},\quad \quad \lambda_j=\mathrm{i} k_j
\label{eq:SC}
\end{equation}

$\mathbf{K}$ and $\mathbf{M}$ are the same as previously, while $\mathbf{L}$ and $\mathbf{H}$ are matrices, functions of $\omega$ and the elasticity tensor \cite{COL11}. A great asset of SCOM is that frequency dependent damping can be taken into account in these matrices. In particular, it is useful for materials with nonlinear frequency dependent damping, like plastics.

As the periodicity is directly included in the structure behaviour equations, the resulting boundary conditions are simple continuity conditions:
\begin{equation}
  \left\{
      \begin{aligned}
		\mathbf{\tilde{u}_{R}}=\mathbf{\tilde{u}_{L}}\\
		\mathbf{\tilde{u}_{U}}=\mathbf{\tilde{u}_{D}}
      \end{aligned}
    \right.
    \label{eq:continuite}
\end{equation}
 
The detailed method can be found in the paper of Collet et al. \cite{COL11}.

\subsection{Used methods}\label{ss:methods}
For both the WFEM and the SCOM, there are two ways to solve the eigenvalue problem:

\begin{description}
	\item[Direct WFE (resp. SCO) method] Here the wavenumber $\lambda_j$ is fixed, and (\ref{eq:WFE}) (resp. (\ref{eq:SC})) is solved to find the eigen pulsations $\omega_j$, and thus the eigenfrequencies $f_j$. The full dispersion diagram is obtained through sweeping $k_x$ and $k_y$ along the IBZ.
	\item[Inverse WFE (resp. SCO) method] Here the frequency, or $\omega$, is fixed, and (\ref{eq:WFE}) (resp. (\ref{eq:SC})) is solved to find the eigenvalues $\lambda_j$, and thus the wavenumbers $k_j$. The full dispersion diagram is obtained through a frequency sweep.
\end{description}

In total, there are four different methods, that we have computed and compared. The comparison is presented in part \ref{ss:methods_comparison}.


\section{Static study}

The first step is the static compression study, which gives the stress-strain curve of the structure. Different states of compression are also used as initial geometries for the dispersion calculations.\\

This study will focus on an infinite plate in the $XY$ plane. Using the Floquet-Bloch theorem mentioned in section \ref{ss:WFEM}, we know that a single periodic pattern is enough to model the whole structure's dynamics, with greatly reduced computation cost. The same pattern will thus be used for statics, but with simple periodic boundary conditions instead. Figs. \ref{fig:cell}b and \ref{fig:cell_enh}b illustrate the concerned patterns. In the $Z$ direction, plane strain approximation is assumed.\\

\subsection{Material} 
The material chosen for the entire study is a silicone exhibiting a very large yield strength to bear large structural deformation without yielding. For the sake of simplicity, a linear elastic material model, necessarily imperfect even though very close to the real material behaviour, is used. \\
\begin{table}[ht!]
  \begin{center}
    \label{tab:silicone_properties}
    \begin{tabular}{|c|c|}
      	\hline
  		\rowcolor{lightskyblue}
      	Material & Silicone\\
      	\hline
      	E & 0.97 MPa\\
      	\hline
      	$\nu$  & 0.499\\
      	\hline
      	$\rho$  & 1150 kg/m$^3$\\
      	\hline
    \end{tabular}
    \caption{Material properties}
  \end{center}
\end{table}

\subsection{Boundary conditions}
A uniaxial compression is considered along $Y$ axis. To that end, a displacement $u_0$ is imposed on the bottom boundary:
$$
u_{\mathrm{B},y}=u_0
$$
Since we model a single pattern, periodic boundary conditions are added:
\begin{itemize}
\item[$\rhd$] $u_{\mathrm{R},y}=u_{\mathrm{L},y}$
\item[$\rhd$] $u_{\mathrm{R},x}=-u_{\mathrm{L},x}$
\item[$\rhd$] $u_{\mathrm{T},x}=u_{\mathrm{B},x}$
\item[$\rhd$] $u_{\mathrm{T},y}=-u_{\mathrm{B},y}$
\end{itemize}
A simple remark is that the last condition doubles the total compression displacement. As a result, $u_0$ amounts to half the effective value, and the effective strain is derived from $\varepsilon={2u_0}/{L_{Py,j}}={u_0}/{L}$.\\
To keep the periodicity orthogonal, the rotation of left boundary is prevented with a homogeneous displacement constraint that can be expressed by ${\partial u_{\mathrm{L},x}}/{\partial y}=0$.

\subsection{Instability}
In numerical simulations, symmetric boundary conditions applied to a symmetric model will always result in a symmetric solution. On the other hand, buckling corresponds to a loss of symmetry in the tangential stiffness matrix. This implies that in our case, the computation will give a purely axial compression, and buckling should not appear.\\
In order to initiate buckling in the structure, an imperfection can be inserted to break the symmetry by a slight amount. To that end, a linear buckling analysis is first performed to obtain the first - and thus most likely - buckling mode. A small fraction (less than 1\pourmille\ of $L_{Px}$ in our case) of this mode displacement field is introduced in the initial structure.\\

\subsection{Results}\label{ss:statics_results}

\begin{figure}[ht!]
\centering
\includegraphics[trim=0cm 0cm 0cm 0.5cm, clip=true, height=0.08\textheight]{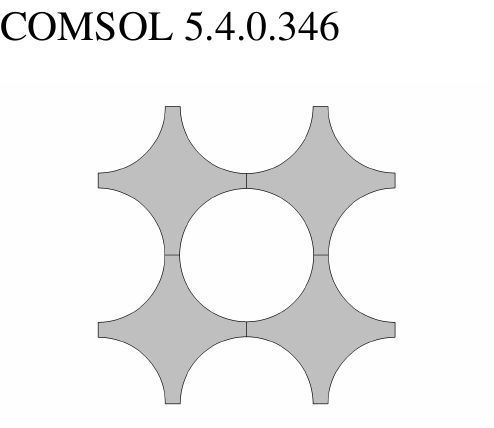}
\hskip 3.7cm 
\includegraphics[trim=0cm 0cm 0cm 0.5cm, clip=true, height=0.08\textheight]{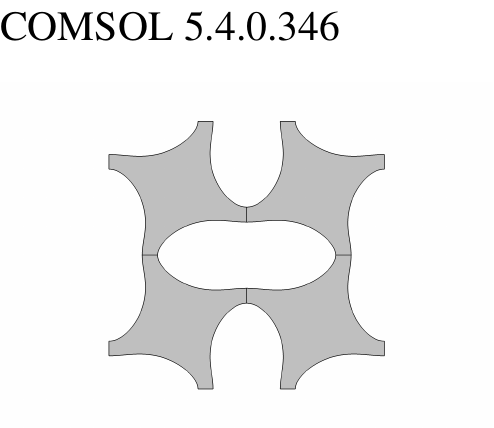}
\\
\includegraphics[width=8.4cm]{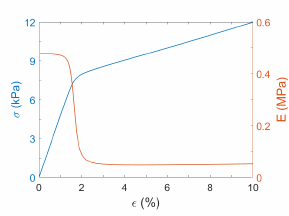}
\caption{Static behaviour of pattern under axial compression. {\color{blue}\protect \textendash} Effective stress $\sigma$ and {\color{red}\protect \textendash} effective Young's modulus $E$ along compression axis ($Y$)}
\label{fig:statics}
\end{figure}

The first simulation to run is the static compression study, which gives the stress-strain curve of the structure. Results for the $S_0$ configuration are shown in figure  \ref{fig:statics}. \\

Even though it may look like a trivial homogeneous yielding material traction curve, the fact that it is a compression curve in elastic domain make it way more peculiar. Two separate parts can be observed: an initial slope relatively stiff compared to second one, flatter. This behaviour is highlighted by the red curve plotting the effective stiffness modulus $E={\Delta\sigma}/{\Delta\varepsilon}$ along the compression axis. Two stable levels are clearly shown, the first one corresponding to a pre-buckling state and the second one associated to post-buckling stiffness.\\

\begin{figure}[ht!]
\begin{minipage}{1.2cm}
	a.
	\vskip 30ex
	b.
	\vskip 20ex
	c.
\end{minipage}
\begin{minipage}{8.4cm}
\includegraphics[width=\linewidth]{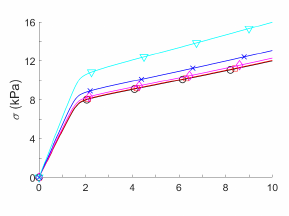}
\\
\includegraphics[width=\linewidth]{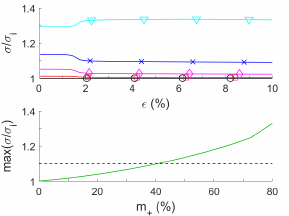}
\end{minipage}

\caption{Radius $R$ effect on static behaviour: (a) effective stress $\sigma$ and (b) normalized effective stress. {\color{black}\protect $\circ$} Initial case, {\color{red}\protect $\Box$} $r=3mm$, {\color{mmagenta}\protect $\diamond$} $r=3.5mm$, {\color{blue}\protect $\times$} $r=4mm$, {\color{mblue}\protect $\bigtriangledown$} $r=4.5mm$. (c) {\color{mgreen}\protect \textendash} Maximum value of normalized effective stress, compared to a $10\%$ threshold ({\color{black}\protect - - -}).}
\label{fig:statics_radius}
\end{figure}

Figure \ref{fig:statics_radius}a shows how modifying the unit cell affects its mechanical behaviour under compression. A few chosen cases illustrate the effect of disk radius on the static stiffness. Insofar as the greatest deformation occurs in the thin ligaments, adding matter around the disk has but small influence on the static results. However, at some point, increasing further the disk radius tends to reduce the thin ligaments length, thus sensibly increasing the effective stiffness.\\
In figure \ref{fig:statics_radius}b, one notices that for the major part of the presented cases, changes in $R$ affects more the pre-buckling stiffness than the post-buckling one, except for the $R=4.5mm$ case. In this configuration, the ligaments are modified to such an extent that the trend derived from the other cases is inverted. Given that we aim to enhance the dynamic properties for unchanged static behaviour, this last case clearly has to be excluded. \\
For the same reason, a threshold can be chosen to determine under which stress increase we can reasonably consider the static behaviour comparable. We arbitrarily chose a $10\%$ limit and plotted in figure \ref{fig:statics_radius}c the maximum stress, normalized by the $S_0$ case stress $\sigma_j$, against the additional mass $m_+=(m-m_{S_0})/{m_{S_0}}$. Compared to the disk radius, $m^+$ is indeed a more sensible variable, as disks with radii under $2.6mm$ bring no change to the initial geometry, and linear radius increase beyond that point entail non-linearly increasing surface - and thus mass - change. The table \ref{tab:radius_masses} summarizes the correspondances between radius and masses.
One can deduce from figure \ref{fig:statics_radius}c that the $10\%$ condition is fulfilled for additional mass under $40\%$. Transposing this for radius values, we know that the condition would be respected for $R\leq3.75mm$. The other cases are nevertheless included for comparison in the dynamic study hereafter.

\begin{table}
\centering
\begin{tabular}{|c|c|c|}
  \hline
  \rowcolor{lightskyblue}
  $R (mm)$ & $m (g)$ & $m_+ (\%)$ \\
  \hline
   $0$ & 6.7 & 0 \\
   $2.75$ & 6.8 & 2.04 \\
   $2.875$ & 7 & 4.57 \\
   $3$ & 7.2 & 7.73 \\
   $3.125$ & 7.5 & 11.46 \\
   $3.25$ & 7.7 & 15.7 \\
   $3.375$ & 8.1 & 20.4 \\
   $3.5$ & 8.4 & 25.55 \\
   $3.625$ & 8.8 & 31.12 \\
   $3.75$ & 9.2 & 37.10 \\
   $3.875$ & 9.6 & 43.46 \\
   $4$ & 10.1 & 50.19 \\
   $4.125$ & 10.5 & 57.28 \\
   $4.25$ & 11 & 64.71 \\
   $4.375$ & 11.5 & 72.48 \\
   $4.5$ & 12.1 & 80.58 \\
  \hline
\end{tabular}
\caption{Correspondences between the disk radius $R$, pattern mass $m$, and additional mass $m_+$}
\label{tab:radius_masses}
\end{table}


\section{Dynamic study}
\subsection{Computation parameters}
The commercial finite elements software COMSOL Multiphysics\textsuperscript{\textregistered} is used in this study. It can automatically generate meshes of various refinements, but for computational cost reduction, some parameters are here manually defined.\\
\begin{figure}[ht!]
\centering
a) \includegraphics[trim=0cm 0cm 0cm 0.5cm, clip=true,width=4.8cm]{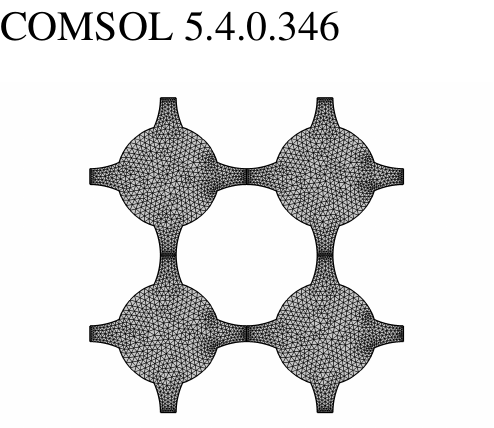}
\hfil
b) \includegraphics[trim=0cm 0cm 0cm 0.5cm, clip=true,width=4.8cm]{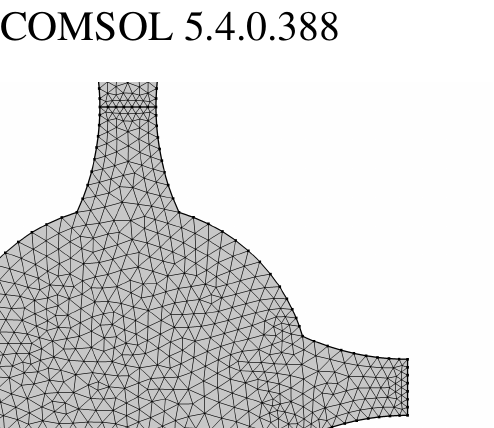}
\caption{Example of generated mesh: (a) periodic pattern and (b) detailed portion}
\label{fig:mesh}
\end{figure}\\
As mentioned before, the greatest strain occurs in the ligaments. Therefore, special 
attention is to be payed when meshing these parts. After a convergence study, it has  been chosen to use seven elements in the thickness of the thinnest part. The mesh is then automatically generated using an advancing front algorithm to tesselate triangular elements which size varies between $1.12\mu m$ and $0.3mm$, with a maximum growth rate of $1.2$ and a curvature factor of $0.25$. An example of resulting mesh is showed in figure \ref{fig:mesh}.\\

One can note that imposed displacement boundary condition implies a finite structure, in opposition to the infinite periodicity assumed in dispersion calculations. However, we suppose the compression to occur very far from the studied pattern, so that side effects can be disregarded.\\

Let finally mention that infinitesimal strain theory being inconsistant here, geometric nonlinearity is taken into account in the computations.\\

\subsection{Calculation method comparison}\label{ss:methods_comparison}

As described in part \ref{ss:methods}, we can choose one method out of four to run our computations. A comparison is thus required to optimize the time and the resolution. To do so, two criteria are showed here: the resolution obtained with a fixed computation time, and the minimum time required to obtain an acceptable resolution on each mode.\\

\begin{figure}[ht!]
\centering
\begin{minipage}{7.2cm}
\includegraphics[width=\linewidth]{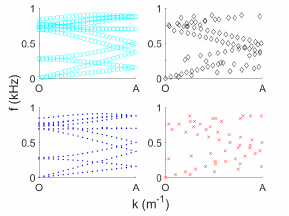}
\end{minipage}
\begin{minipage}[t]{1.8cm}
\includegraphics[trim=0cm 0cm 0cm 0cm, clip=true, width=\linewidth]{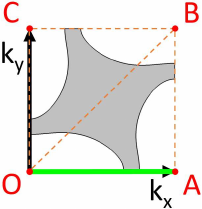}
\end{minipage}
\caption{Band diagram for OA branch of IBZ, considering the $S_0$ pattern under a $10\%$ axial compression along $Y$ axis. {\color{mblue}\protect $\circ$} WFEM (top left), {\color{blue}\protect $\cdot$} SCOM (bottom left), {\color{black}\protect $\diamond$} inverse WFEM (top right), and {\color{red}\protect $\times$} inverse SCOM (bottom right).}
\label{fig:comp_method}
\end{figure}

For the first comparison, a fixed computation time is chosen (90 seconds in our case) and the resulting diagrams are plotted in figure \ref{fig:comp_method}. For the sake of clearness, only the modes under $1kHz$ for the OA branch of the IBZ are presented.\\

A clear difference is observed between direct methods (cyan circles and blue dots) and inverse ones (black diamonds and red crosses). On the former, where the wavenumber $k$ is fixed and where the eigenfrequencies are found, the eigenmodes are very easy to identify and read, while on the latter, where $f$ is fixed to search for the wavenumbers, there are fewer points on the mode branches and the reading is more troublesome, in particular for almost flat branches, i.e. for slow group velocities. The mode around $700Hz$ is for example invisible to inverse methods, unless running calculations with extremely fine sweep resolutions, which would greatly increase the computational cost. Among direct methods, the WFEM (27 points per mode) gives slightly better results than SCOM (21 points per mode), but the gap increases for inverse methods, where the WFEM (resp. SCOM) gives a resolution of $20Hz$ (resp. $38Hz$). An advantage of inverse methods is that usually fewer eigenvalues are needed than with direct ones, but as it can be seen, the time gain is far from enough to compensate for the resolution loss. 

\begin{figure}[ht!]
\centering
\begin{minipage}{5.4cm}
a)\includegraphics[width=\linewidth]{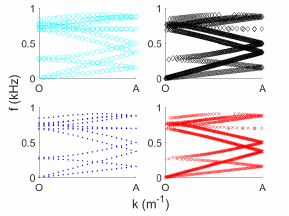}
\end{minipage}
\hfil
\begin{minipage}{5.4cm}
b)\includegraphics[width=\linewidth]{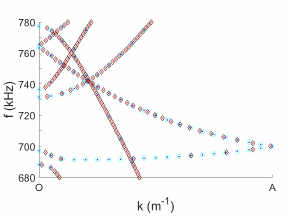}
\end{minipage}
\caption{Band diagram for OA branch of IBZ, considering the $S_0$ pattern under a $10\%$ axial compression along $Y$ axis. {\color{mblue}\protect $\circ$} WFEM, {\color{blue}\protect $\cdot$} SCOM, {\color{black}\protect $\diamond$} inverse WFEM, and {\color{red}\protect $\times$} inverse SCOM. a) Diagrams for each method and b) comparison for modes between $680Hz$ and $780Hz$.}
\label{fig:comp_method_res}
\end{figure}

An appreciation of the computational cost difference can be seen through the second criterion. As perfectly identical resolutions between inverse and direct methods cannot be obtained, we compare the time required to obtain an acceptable resolution on each mode. The corresponding diagrams are shown in figure \ref{fig:comp_method_res}. 

Direct methods are very easy to tune, as the number of points on each mode is the number of steps input in the study. Here we chose 21 points along $k$ to obtain the cyan and blue curves, where each branch is easy to follow. The results are exactly the same, but the WFEM is slightly faster (1'26) than the SCOM (1'29).\\
On the contrary, results with inverse methods (in black and red) are very sensitive to group velocities; the flatter the modes, the finer the resolution must be. Here we chose a $2Hz$ resolution, which took 18'15 for WFEM and 25'20 for SCOM. Despite the great cost, the closeup in figure \ref{fig:comp_method_res}b shows that the flat mode below $700Hz$ remains hard to read.

Faster and easier to read, direct methods thus seem better suited for our computations. As mentioned in the SCOM description, a major advantage of the latter is that it can take into account the damping in the structure, unlike the WFEM. However, this parameter being neglected in this study, the direct WFE method is preferred, as it is slighlty faster than its SCOM counterpart.

\subsection{Internal stress effect}

Dispersion calculations are based on a periodic pattern geometry to determine the propagation modes in the structure. But even though the evolution of these modes can be observed as function of parameters like the effective strain rate, the starting point of each dynamic computation is, by default, the deformed geometry without the stress caused by the static compression. It seems interesting to check the actual effect of such internal stress, in particular since the strain levels can be high (up to $40\%$ in this study range). In the software used, this requires to extract the stress field obtained on the static compression, and add it as initial condition to the deformed geometry for dispersion calculation.

\begin{figure}[ht!]
\centering
a)\includegraphics[width=7.2cm]{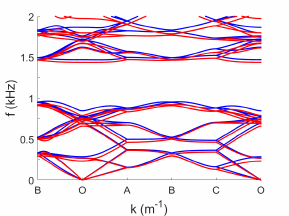}\\
b)\includegraphics[width=7.2cm]{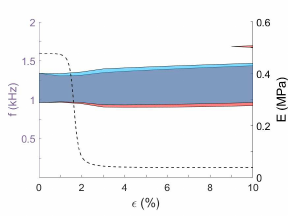}
\caption{Internal stress effect. a) Band diagram of $S_0$ pattern under $Y$-axis $10\%$ compression, with WFEM. {\color{blue}\protect \hdashrule[0.5ex]{3mm}{1pt}{}} Unstressed configuration and {\color{red}\protect \hdashrule[0.5ex]{3mm}{1pt}{}} with internal stress. b) Evolution of omnidirectional bandgaps range under $2kHz$ as function of deformation: \protect\tikz \protect\shade [shading=sblues] (1,-1) rectangle (1.2,-1.2);\ Unstressed configuration bandgap,  \protect\tikz \protect\shade [shading=reds] (0,0) rectangle (0.2,0.2);\ prestressed configuration bandgap, and \hdashrule[0.5ex]{7mm}{1pt}{3pt}effective Young's modulus along compression axis}
\label{fig:stress_effect}
\end{figure}

The effect of internal stress is illustrated in figure \ref{fig:stress_effect}. It can be seen that stress tends to soften the structure, as the modes frequencies in figure \ref{fig:stress_effect}a are shifted downwards.\\

As figure \ref{fig:stress_effect}b shows, internal stress affects nonlinearly the first bandgap range. While the lower boundary mainly downshifts for buckled configurations, the upper one is already shifted for $\varepsilon=1\%$ and barely increase after that. For both, the shift stabilizes after a few percent of deformation.\\
Finally a second bandgap can be seen in red around $1.7kHz$, which was not detected during the unstressed calculations.\\

All this shows that not only the bandgap width but also their number can be incorrect if internal stress is not implemented. Since it seems closer to real case, internal stress are thus taken into account in hereinafter results.
 
\subsection{Geometric optimization}
After validating simulation parameters with previous steps, comparison between initial pattern $S_0$ and modified ones (Figure \ref{fig:cell_enh}) is performed. Computations are run with direct WFEM for all cases in table \ref{tab:radius_masses} and for precompression states between $0\%$ and $10\%$ with a $1\%$ step.\\
Three different criteria are finally used to evaluate the structures' dynamic performance: the first bandgap width $w_{\mathrm{BG_1}}$, the frequency range covered by all the bandgaps under $2kHz$ $w_{\mathrm{BG,t}}$, and the lowest inferior bound of the first bandgap $f_{\mathrm{L,BG_1}}$, compared to initial case $S_0$.\\

\begin{figure}[ht!]
\centering

\begin{minipage}{4.92cm}
\begin{minipage}{\linewidth}
	\centering
	\vskip 5ex
	First geometry
\end{minipage}
\begin{minipage}{0.46\linewidth}
	\centering
	\vskip 2ex
	$S_0$
\end{minipage}
\begin{minipage}{0.52\linewidth}
	\centering
	\vskip 2ex
	\includegraphics[trim=0cm 0cm 0cm 0.5cm, clip=true, width=\linewidth]{figure_embedded_6_12t_12l_13t_13l-eps-converted-to.pdf}
\end{minipage}
\vskip 14ex
\begin{minipage}{\linewidth}
	\centering
	Second geometry
\end{minipage}
\begin{minipage}{0.46\linewidth}
	\centering
	\vskip 2ex
	$R=3.25mm$
\end{minipage}
\begin{minipage}{0.52\linewidth}
	\centering
	\vskip 2ex
	\includegraphics[trim=0cm 0cm 0cm 0.5cm, clip=true, width=\linewidth]{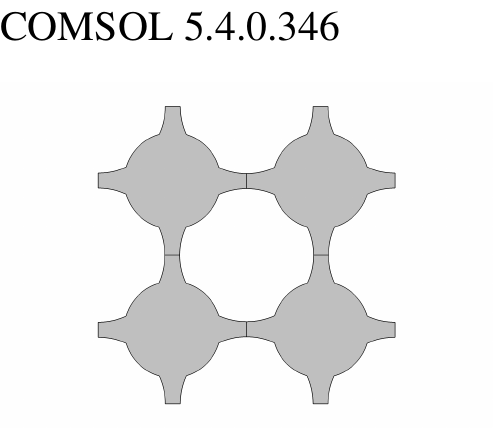}
\end{minipage}
\vskip 12ex
\begin{minipage}{\linewidth}
	\centering
	Third geometry
\end{minipage}
\begin{minipage}{0.46\linewidth}
	\centering
	\vskip 2ex
	$R=4.25mm$
\end{minipage}
\begin{minipage}{0.52\linewidth}
	\centering
	\vskip 2ex
	\includegraphics[trim=0cm 0cm 0cm 0.5cm, clip=true, width=\linewidth]{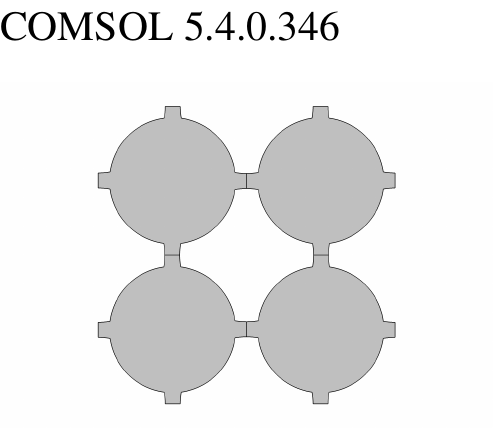}
\end{minipage}
\end{minipage}
\begin{minipage}{6.48cm}
	\includegraphics[trim=0cm 0cm 0cm 0.5cm, clip=true, width=0.35\linewidth]{figure_embedded_6_12t_12l_13t_13l-eps-converted-to.pdf}
	\hfill
	\includegraphics[trim=0cm 0cm 0cm 0.5cm, clip=true, width=0.35\linewidth]{figure_embedded_6_12r_13b-eps-converted-to.pdf}\\
	\includegraphics[trim=0cm 0.2cm 0cm 0cm, clip=true, width=\linewidth]{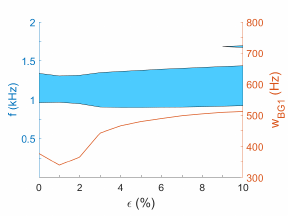}
	\includegraphics[trim=0cm 0.2cm 0cm 0cm, clip=true, width=\linewidth]{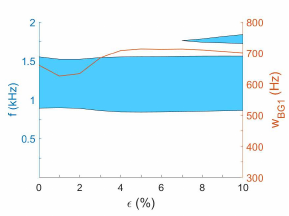}
	\includegraphics[width=\linewidth]{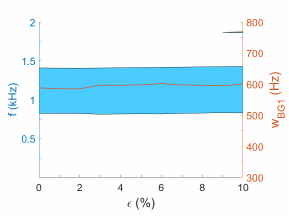}
\end{minipage}
\caption{Evolution of omnidirectional bandgap ranges as function of deformation for three chosen geometries. \protect\tikz \protect\shade [shading=sblues] (1,-1) rectangle (1.2,-1.2);\ Bandgaps and {\color{red}\protect \hdashrule[0.5ex]{3mm}{1pt}{}} width $w_{\mathrm{BG_1}}$ of the first bandgap}. 
\label{fig:strain_effect_radius}
\end{figure}

Figure \ref{fig:strain_effect_radius} depicts the evolution of omnidirectional bandgap ranges below $2kHz$ as function of static deformation. While the deformation has a great effect on the first bandgap of $S_0$ geometry, particularly for $\varepsilon<5\%$, this effects seems to be fading for larger values of $R$. The $w_{\mathrm{BG_1}}$ curve highlights very well this effect, as it is getting flatter with large $R$.\\

The bandgap narrowing for compression smaller than $2\%$ may be explained partly by the stress increase in the material, and partly by the axial length reduction. Since the structure is not buckled at that point, the compression mode is globally longitudinal, so the internal stress rapidly increases. Combined to the $L_y$ decrease, it results in the bandgap narrowing observed. After $2\%$, the buckling induced stiffness drop causes the bandgaps to widen.\\

\begin{figure}[ht!]
\centering
\begin{minipage}{1.44cm}
	\centering
	\vskip 8ex
	$\varepsilon=0\%$
	\vskip 27ex
	$\varepsilon=2\%$
	\vskip 27ex
	$\varepsilon=10\%$
\end{minipage}
\begin{minipage}{3cm}
	\centering
	\vskip 8ex
	\includegraphics[trim=0cm 0cm 0cm 0.5cm, clip=true, width=\linewidth]{figure_embedded_6_12t_12l_13t_13l-eps-converted-to.pdf}
	\vskip 15ex
	\includegraphics[trim=0cm 0cm 0cm 0.5cm, clip=true, width=\linewidth]{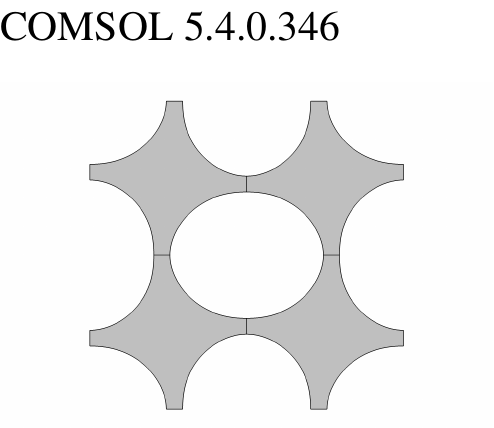}
	\vskip 15ex
	\includegraphics[trim=0cm 0cm 0cm 0.5cm, clip=true, width=\linewidth]{figure_embedded_6_12r_13b-eps-converted-to.pdf}
\end{minipage}
\begin{minipage}{7.2cm}
	\includegraphics[trim=0cm 0cm 0cm 0.5cm, clip=true, width=0.3\linewidth]{figure_embedded_6_12t_12l_13t_13l-eps-converted-to.pdf}
	\hspace{2cm}
	\includegraphics[trim=0cm 0cm 0cm 0.5cm, clip=true, width=0.3\linewidth]{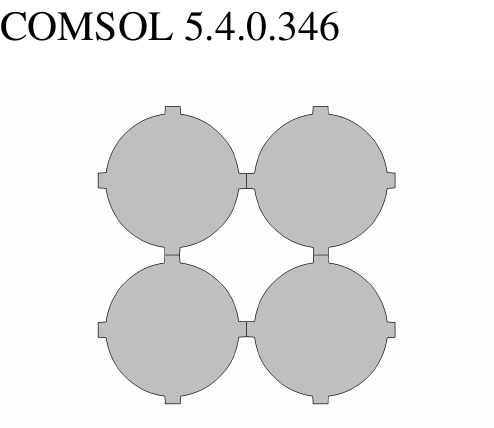}\\
	\includegraphics[trim=0cm 0.2cm 0cm 0cm, clip=true, width=0.9\linewidth]{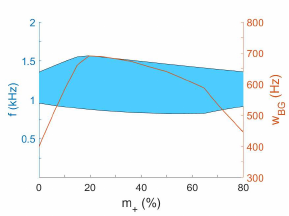}
	\includegraphics[trim=0cm 0.2cm 0cm 0cm, clip=true, width=0.9\linewidth]{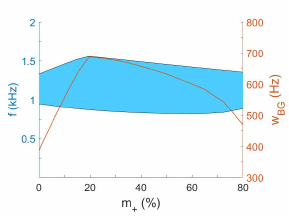}
	\includegraphics[width=0.9\linewidth]{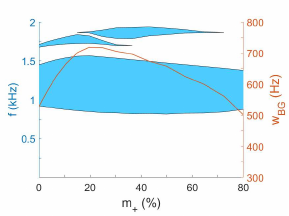}
\end{minipage}
\caption{Evolution of omnidirectional bandgap ranges as function of additional mass for three compression levels. \protect\tikz \protect\shade [shading=sblues] (1,-1) rectangle (1.2,-1.2);\ Bandgaps and {\color{red}\protect \hdashrule[0.5ex]{3mm}{1pt}{}} width $w_{\mathrm{BG_1}}$ of the first bandgap.}
\label{fig:mass_effect_strain}
\end{figure}

Another point is that the bandgaps size for the second geometry appears to be much greater than for $S_0$. The difference is particularly obvious in figure \ref{fig:strain_effect_radius} for low $\varepsilon$ values, as well as for the second bandgap, around $1.75kHz$. One could think that increasing $R$ even further could enhance this effect but the opposite is observed: bandgaps are thinner for $R=4.25mm$. The analysis of the $w_{\mathrm{BG_1}}$ curves confirms that the width increases for $R=3.25mm$ compared to $S_0$, but then decreases for $R=4.25mm$. Therefore plotting the bandgap width as function of the radius would allow to check if an optimum exists. These plots can be seen in figure \ref{fig:mass_effect_strain}, but as function of additional mass for the same reason as mentioned in the static results section.\\

\begin{figure}[ht!]
	\centering
	\begin{tikzpicture}
		\node[anchor=south west,inner sep=0] (image) at (0,0,0){\includegraphics[width=7.2cm]{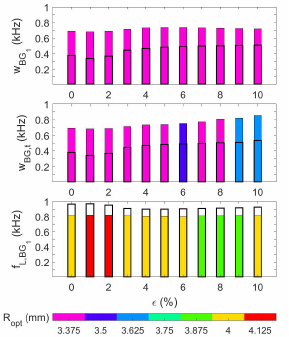}};
		\begin{scope}[x={(image.south east)},y={(image.north west)}]
			\draw(0.24,0.059) node {$\circ$};
			\draw(0.35,0.059) node {$\diamond$};
			\draw(0.46,0.059) node {$\vartriangle$};			
			\draw(0.57,0.059) node {$\times$};
			\draw(0.68,0.059) node {$\star$};
			\draw(0.79,0.059) node {$\ast$};
			\draw(0.90,0.059) node {$\bullet$};
			
			\draw(0.247,0.892) node {$\circ$};
			\draw(0.311,0.89) node {$\circ$};
			\draw(0.376,0.893) node {$\circ$};
			\draw(0.440,0.899) node {$\circ$};
			\draw(0.505,0.903) node {$\circ$};
			\draw(0.569,0.904) node {$\circ$};
			\draw(0.633,0.904) node {$\circ$};
			\draw(0.697,0.902) node {$\circ$};
			\draw(0.761,0.901) node {$\circ$};
			\draw(0.8265,0.90) node {$\circ$};
			\draw(0.891,0.899) node {$\circ$};
					
			\draw(0.247,0.604) node {$\circ$};
			\draw(0.311,0.602) node {$\circ$};
			\draw(0.376,0.603) node {$\circ$};
			\draw(0.440,0.609) node {$\circ$};
			\draw(0.505,0.614) node {$\circ$};
			\draw(0.569,0.618) node {$\circ$};
			\draw(0.633,0.619) node {$\diamond$};
			\draw(0.697,0.623) node {$\circ$};
			\draw(0.761,0.631) node {$\circ$};
			\draw(0.8265,0.635) node {$\vartriangle$};
			\draw(0.891,0.642) node {$\vartriangle$};
					
			\draw(0.247,0.345) node {$\ast$};
			\draw(0.311,0.347) node {$\bullet$};
			\draw(0.376,0.347) node {$\bullet$};
			\draw(0.440,0.345) node {$\ast$};
			\draw(0.505,0.344) node {$\ast$};
			\draw(0.569,0.344) node {$\ast$};
			\draw(0.633,0.343) node {$\ast$};
			\draw(0.697,0.345) node {$\star$};
			\draw(0.761,0.345) node {$\star$};
			\draw(0.8265,0.345) node {$\star$};
			\draw(0.891,0.345) node {$\ast$};
		\end{scope}
	\end{tikzpicture}
			
	\caption{Optimization results for three criteria: $w_\mathrm{BG_1}$ (top), $w_\mathrm{BG,t}$ (center), and $f_\mathrm{L,BG_1}$ (bottom). \protect\tikz \protect\shade [shading=rainbow, shading angle=180] (1,-1) rectangle (1.2,-1.2); Best result for each deformation compared to $\Box$ reference case $S_0$.}
	\label{fig:optim_res}
\end{figure}

Figure \ref{fig:mass_effect_strain} shows indeed a non monotone variation of bandgaps, whether it be regarding their lower boundary, upper boundary, or their width. The optimum case may vary depending on which of these criteria is chosen, but the mass (or radius) corresponding to this optimum seems to be bearly affected by the observed deformation. Considering the mass $m_+$ dependence, the $w_{\mathrm{BG_1}}$ curves display quite flat optimum zones, which is a good asset for the robustness, as small variations of mass will not affect much the results. Figure \ref{fig:optim_res} and table \ref{tab:optima} summarize the optimization results for each state of deformation and for three different criteria, namely the first bandgap width $w_\mathrm{BG_1}$, the frequency range covered by all the bandgaps under $2kHz$ $w_\mathrm{BG,t}$, and the lowest inferior bound of the first bandgap $f_\mathrm{L,BG_1}$, compared to initial cas $S_0$.\\

The first criterion (Figure \ref{fig:optim_res} - top) is quite straightforward, since the optimal radius is the same for all the deformation states, namely $R_\mathrm{opt}=3.375mm$. Even though $\varepsilon$ has but little influence on the optimal values, an optimal precompression is found at $5\%$, at which point the first bandgap is the widest, with $w_\mathrm{BG_1}=734.5Hz$. We note a huge improvement $w_\mathrm{BG_1}$ in particular for small precompressions, with up to $101\%$ bandgap width increase for $\varepsilon=1\%$.\\

For deformations lower than $5\%$, there is only one omnidirectional bandgap below $2kHz$ so the $w_\mathrm{BG,t}$ optimization (Figure \ref{fig:optim_res} - center) gives the exact same result as for $w_\mathrm{BG_1}$. However for $\varepsilon\geq6\%$, one or two smaller bandgaps open and as the figure \ref{fig:optim_res} shows, the total frequency range covered by bandgaps increases, even though the first bandgap is narrowing. Moreover the optimum radius tends to be higher than before, with $R_\mathrm{opt}=3.625mm$ for the two greatest values of $w_\mathrm{BG,t}$. We could probably find an optimal value for greater levels of $\varepsilon$, but this study range does not give such a result. In the end, the optimization study is quite efficient for $w_\mathrm{BG,t}$, with an improvement between $53\%$ and $101\%$. Finally, contrary to the other criteria, we can see that the deformation tends to improve significantly $w_\mathrm{BG,t}$, with a $38\%$ increase between the optimal values of $\varepsilon=10\%$ compared to the unloaded case. \\

The third criterion (Figure \ref{fig:optim_res} - bottom) gives sensibly different results, as the optimum radius is higher than before ($R_\mathrm{opt}\approx4mm$) and the enhancement due to deformation is quite weak. The lowest bandgap frequency achieved is $799Hz$ for $R_\mathrm{opt}=4mm$ at $\varepsilon=6\%$, only $18Hz$ lower than the same geometry at zero compression. The geometry effect is nonetheless important as a $15\%$ frequency drop is observed between the optimal case and $S_0$ for the first two deformations. If we add the static threshold discussed in section \ref{ss:statics_results}, this value decreases to $13\%$ at $R_\mathrm{opt}=3.75mm$, which is still substantial. 

\begin{table}[ht!]
\centering
\begin{tabular}{|c||c|c|c|c|c|c|c|c|c|}
  \hline
   	& \multicolumn{3}{c|}{Widest $\mathrm{BG_1}$}  & \multicolumn{3}{c|}{Total BGs range} & \multicolumn{3}{c|}{$\mathrm{BG_1}$ lowest bound} \\
  \cline{2-10}
  & $R$ (mm) & \multicolumn{2}{c|}{$w_\mathrm{BG_1}$ (Hz)} & $R$ (mm) & \multicolumn{2}{c|}{$w_\mathrm{BG,t}$ (Hz)} & $R$ (mm) & \multicolumn{2}{c|}{$f_\mathrm{L,BG_1}$ (Hz)}\\
  \cline{2-10}
  \multirow{-3}*{$\varepsilon$ (\%)}& Opt & Opt & $S_0$ & Opt & Opt & $S_0$ & Opt & Opt & $S_0$\\
  \hline
   $0$ 	& 						 	& 690 & 377 &							& 690 & 377 & $4$ 						& 817 & 962 \\
   \cdashline{8-8}
   $1$ 	& 							& 682 & 339 &							& 682 & 339 & 		 					& 821 & 967 \\
   $2$ 	& 							& 687 & 366 &							& 687 & 366 & \multirow{-2}*{$4.125$} 	& 820 & 949 \\
   \cdashline{8-8}
   $3$ 	& 							& 713 & 443 &							& 713 & 443 & 							& 811 & 906 \\
   $4$ 	& 							& 731 & 466 &							& 731 & 466 & 							& 803 & 898 \\
   $5$ 	& 							& 734 & 479 & \multirow{-6}*{$3.375$}	& 734 & 479 & 							& 800 & 898 \\
   \cdashline{5-5}
   $6$ 	& 							& 734 & 489 & $3.5$ 					& 748 & 489 & \multirow{-4}*{$4$}		& 799 & 903 \\
   \cdashline{5-5}
   \cdashline{8-8}
   $7$ 	& 							& 731 & 498 &							& 772 & 498 & 							& 806 & 905 \\
   $8$ 	& 							& 728 &	504 & \multirow{-2}*{$3.375$} 	& 807 & 504 & 							& 810 & 911 \\
   \cdashline{5-5}
   $9$ 	& 							& 724 & 509 &					 		& 821 & 509 & \multirow{-3}*{$3.875$}	& 814 & 917 \\
   \cdashline{8-8}
   $10$ & \multirow{-11}*{$3.375$} 	& 719 & 511 & \multirow{-2}*{$3.625$} 	& 853 & 534 & $4$						& 814 & 924 \\
  \hline
\end{tabular}
\caption{Optimization results for different criteria}
\label{tab:optima}
\end{table}


\section{Conclusion}
In the end, static and dynamic computations have been run for both a reference case and modified geometries. The static study has shown that the reference geometry can be modified up to a certain extent without much of a change in the effective stress-strain curve for axial compression. For all cases, the static curve exhibit a deep stiffness drop triggered by buckling that could be used to mechanically tune the structure properties.\\

Given that such compression entails inhomogeneous stress field, the effect of the latter has been demonstrated through comparison with unstressed calculation, which has seldom been observed in litterature. The position, width, and number of bandgaps is affected by the stress field. Using this result, and after comparing different dispersion calculation methods, an optimization study has been performed on an original design, inspired from classical rotating squares geometry. As a result, the dynamic properties of the structure can be greatly improved by such geometrical modifications. While the precompression has a certain effect on bandgaps, the unit cell modification can result in huge improvements on bandgaps widths and position.\\

As prospects, the bandgaps analysis might be validated by finer sweep in the Irreducible Brillouin Zone, and completed with unidirectional bandgap analysis, as well as an extended calculation for three-dimensional propagation. The stress levels inside the material could also be compared to failure stress, in particular fatigue failure stress. An important prospect is the finite structure computations along with experimental validation, as boundary effects can greatly modify the whole structure behaviour.

\section*{Acknowledgement}
This work has been carried out thanks to a CIFRE grant with PSA Group company.
The research leading to these results has been performed under the collaborative Framework OpenLab PSA Automobiles Vibro-Acoustic-Tribology@Lyon. The authors gratefully acknowledge them for supporting this program. 

\section*{References}

\bibliography{references}
\end{document}